\documentclass[debug,preprint,cite,overfull]{epl}

\usepackage{amsmath,amssymb}

\usepackage{bm}
\usepackage{epsfig}
\usepackage{psfrag}

\newcommand{\bd}{\bm}

\title{Spin-wave interactions in  quantum antiferromagnets} 
\author{Nils Hasselmann \and Peter Kopietz}
\shortauthor{N. Hasselmann \etal}
\institute{
   Institut f\"{u}r Theoretische Physik, Universit\"{a}t
  Frankfurt,  Max-von-Laue Strasse 1, \\ D-60438 Frankfurt, Germany
}


\begin{document}

\maketitle

\begin{abstract}
We study spin-wave interactions in quantum antiferromagnets
by expressing the usual
magnon annihilation and creation operators 
in terms of Hermitian field operators 
representing transverse staggered and ferromagnetic  spin fluctuations.
In this parameterization, which was anticipated by Anderson in 1952, 
the two-body interaction vertex between staggered spin fluctuations 
vanishes at long wavelengths.
We derive a new effective action for the staggered fluctuations only
by tracing out the  ferromagnetic fluctuations.
To one loop order, the renormalization group flow agrees with the
nonlinear-$\sigma$-model approach.
\end{abstract}

The spin-wave approach to quantum Heisenberg 
antiferromagnets, which was pioneered
by Anderson \cite{Anderson52} and Kubo \cite{Kubo52} more than half a century ago,
is still one of the most powerful methods to calculate the low-temperature properties
of ordered magnets.
In this approach, the components of the quantum-mechanical spin-operators
$\bd{S}_i$ are expressed in terms of canonical boson operators $b_i$ using the
Holstein-Primakoff \cite{Holstein40} or the Dyson-Maleev~\cite{Dyson56} 
transformation.
The spin-$S$ antiferromagnetic Heisenberg  Hamiltonian

\begin{equation}
  \hat{H} =  
\frac{1}{2} 
\sum_{ij} J_{ij} 
\bd{S}_i  \cdot \bd{S}_j  
  \; ,
  \label{eq:hamiltonian}
\end{equation}
can then be written as a bosonic many-body Hamiltonian
$ \hat{H} = - D N J S^2 + \hat{H}_2 + \hat{H}_{\rm int}$,
where the quadratic part is
 \begin{equation}
 \hat{H}_2 = S \sum_{ij} J_{ij}  
[ b^{\dagger}_i b_i +  b^{\dagger}_j b_j
 + b_i b_j + b^{\dagger}_i b^{\dagger}_j ] 
\; . 
 \label{eq:H2}
\end{equation}
The spins are assumed here to be
localized at the sites $\bd{r}_i$ of a
$D$-dimensional hypercubic lattice with
lattice spacing $a$. 
The lattice is bipartite, with sublattices labelled $A$ and $B$.
We also assume
$J_{ij} = J > 0$ for all pairs of nearest neighbors and $J_{ij}=0$ otherwise.

The interaction  $\hat{H}_{\rm int}$ involves 
at  least four boson operators and
higher powers of 
the small parameter $1/S$, so that for large $S$
it is reasonable to treat 
$\hat{H}_{\rm int}$ perturbatively.
However, as we discuss in more detail below, in momentum space the vertices
of $\hat{H}_{\rm int}$
have a complicated non-analytic structure
for small momenta \cite{Harris71,Kopietz90,Igarashi},
so that the expected  supression of the interaction between 
long-wavelength Goldstone modes~\cite{Hofmann99}  is not manifest
in this approach.
On the other hand, 
in the calculation of physical quantities that are dominated by
staggered spin-fluctuations the leading momentum-dependence
of the vertices eventually cancels \cite{Maleev00}.
Note that the $1/S$-expansion for quantum 
ferromagnets is not plagued by this problem: in this case
the two-body interaction between ferromagnetic magnons
vanishes  quadratically for small momenta, reflecting the supression
of the effective interaction between long-wavelength
Goldstone-modes \cite{Kopietz89}.

Alternatively, we may work with 
the nonlinear sigma model (NLSM), which is
believed to describe the long-wavelength and low-energy physics
of quantum Heisenberg antiferromagnets in the
so-called renormalized classical regime \cite{Chakravarty89}
(for a careful discussion and a derivation of an effective model
for the short wavelength regime see
\cite{Belinicher02}).
In imaginary time $\tau$ and at finite temperature $T = 1/\beta$ the action of the NLSM is
 \begin{eqnarray}
S_{\rm{NLSM} }
[ \bd{\Omega} ]  & = & \frac{ \rho_0}{2} \int_0^{\beta}\hspace{-2mm} d \tau
\hspace{-1mm} \int \hspace{-1mm} d^{D} r  \Bigl[ 
( \partial_{\mu} \bd{\Omega})^2 
 + c_0^{-2} ( \partial_{\tau}  \bd{\Omega})^2 \Bigr]
,\hspace{2mm}
 \label{eq:sigma} 
 \end{eqnarray}
where the unit vector  $\bd{\Omega} (  \bd{r} , \tau)$ 
represents the slowly  fluctuating staggered magnetization,
$\rho_0$  and $c_0$ are the  spin stiffness and the spin-wave velocity
at $T =0$. Assuming local staggered order
in  $z$-direction, 
we  may resolve the
constraint  $\bd{\Omega}^2 = 1$ by writing the $z$-component
of    $\bd{\Omega}$ as
$ \sqrt{1-{\bd{\Pi}}^2} \approx 1 - \frac{1}{2} \bd{\Pi}^2 + \ldots$,
where the two-component vector
$\bd \Pi$ represents the transverse staggered spin
fluctuations.
The interaction vertices generated in Eq.~(\ref{eq:sigma})
by expanding the square root 
involve two derivatives,
so that their Fourier transform vanishes quadratically 
for small wave-vectors or frequencies, in contrast to the
vertices in the usual $1/S$-expansion \cite{Harris71,Kopietz90}.

Whether it is possible to parameterize the $1/S$-expansion such
that the weakness of the interaction between long-wavelength staggered 
spin fluctuations
is manifest is a long-standing unsolved problem in spin-wave theory, 
which we shall
solve in this work.
Let us therefore recall the usual diagonalization procedure
of the quadratic spin-wave Hamiltonian (\ref{eq:H2}).
First of all, we introduce the Fourier components of the
operators $b_i$ in the sublattice basis, defining 
$b_i = (2/N)^{1/2}
      \sum_{ \bd{k} } e^{ i \bd{k} \cdot
        \bd{r}_i } A_{\bd{k}}$ if $\bd{r}_i$ belongs to sublattice A, and
$b_i = (2/N)^{1/2}
      \sum_{ \bd{k} } e^{ i \bd{k} \cdot
        \bd{r}_i } B_{\bd{k}}$ if  $\bd{r}_i$ belongs to sublattice B.
Here and below the $\bd{k}$-sums are over the reduced Brillouin zone.
The complete diagonalization of $\hat{H}_2$ is then achieved
with the help of a Bogoliubov transformation,
\begin{equation}
  \left( \begin{array}{c}
      A_{ \bd{k} } \\
      B^{\dagger}_{ - \bd{k}  }  \end{array}
  \right) =
  \left( \begin{array}{cc}
      u_{ \bd{k} } & -  v_{\bd{k} } \\
      -  v_{\bd{k} } & u_{ \bd{k} } \end{array} \right)
  \left( \begin{array}{c}
      \alpha_{ \bd{k}  } \\
      \beta^{\dagger}_{ - \bd{k}  }  \end{array}
  \right)
  \; ,
  \label{eq:bogoliubov}
\end{equation}
where $u_{ \bd{k} } =  2^{-1/2}[\epsilon_{\bd{k}}^{-1}+1]^{1/2}$
and $v_{ \bd{k} } =  2^{-1/2}[\epsilon_{\bd{k}}^{-1}-1]^{1/2}$ with
$\epsilon_{ \bd{k}}    =   [
  1 - \gamma_{\bd{k}}^2 ]^{1/2}$.
Here $\gamma_{\bd k}= D^{-1} \sum_\mu
\cos(   k_\mu  a  )$, where 
$k_\mu = \hat{\bd{e}}_\mu \cdot    {\bd k} $ are the components
of $\bd{k}$ in the direction of the unit vectors
$\hat{\bd e}_\mu$, $\mu =1, \ldots, D$.  
Then we obtain $\hat{H}_2 = - N D JS + \hat{H}_2^{\prime}$, where
\begin{equation}
 \hat{H}_{2}^{\prime} =  \sum_{\bd k} E_{\bd k} \bigl[ \alpha_{\bd k}^\dagger \alpha_{\bd k}
 +\beta_{\bd k}^\dagger \beta_{\bd k} + 1
 \bigr] \; , 
 \end{equation}
with the magnon dispersion $E_{\bd{k}} = 2 D J S \epsilon_{\bd{k}}$.
The Bogoliubov transformation (\ref{eq:bogoliubov}) is not unique: any rotation
that mixes the operators $\alpha_{\bd{k}}$ and $\beta_{\bd{k}}$ will also diagonalize
$\hat{H}_2$. To make contact with the $\bd{\Pi}$-field in the NLSM, 
we choose $\hat{\Psi}_{ \bd{k} \sigma} = 2^{-1/2} ( \alpha_{\bd{k}} + \sigma 
 \beta_{\bd{k}} )$, where $\sigma = \pm $ labels the symmetric and antisymmetric
combinations.
We then express the 
magnon annihilation operators $\hat{\Psi}_{  \bd{k} \sigma }$
in terms of two Hermitian field operators
$\hat{\Pi}_{\bd{k} \sigma}$ and $\hat{\Phi}_{\bd{k} \sigma}$
as follows,
  \begin{eqnarray}
    \hat{\Psi}_{ \bd{k} \sigma} & = & p_{\sigma} 
    ( \chi_0/  2 V  E_{\bd{k}} )^{1/2}
    [ 
      E_{\bd{k}} \hat{\Pi}_{ \bd{k} \sigma} 
      + i\chi_0^{-1} \hat{\Phi}_{ \bd{k}  \sigma} ]
    \label{eq:psiplus}
    \; , \hspace{7mm}
  \end{eqnarray}
where $\chi_0 = ( 4 D J a^D )^{-1} = \rho_0 / c_0^2 $ is the classical uniform 
transverse susceptibility.  The phase factors
$p_{ +} = -i$ and $p_{-} =1$ are introduced
for later convenience and $V = a^D N$ is the volume.
One easily verifies that
 $ [ \hat{\Pi}_{\bd{k} \sigma } , \hat{\Phi}_{\bd{k}^{\prime} \sigma^{\prime}} ]
= i V \delta_{ \bd{k} , - \bd{k}^{\prime} } \delta_{ \sigma , \sigma^{\prime}}$,
so that
 $\hat{\Pi}_{\bd{k} \sigma}$ and 
$\hat{\Phi}_{\bd{k} \sigma}$ are 
canonically conjugate bosonic field operators, with
$ \hat{\Pi}_{\bd{k} \sigma }$/$\hat{\Phi}_{ \bd{k} \sigma}$  
corresponding to the position/momentum operators.
These operators are simply related to
the Fourier components of the
total spin $\bd{S}_{\bd{k}} = N^{-1/2} \sum_i e^{ - i \bd{k} \cdot \bd{r}_i }
 \bd{S}_i$ and the staggered spin
 $\bd{S}_{{\rm st}, \bd{k}} = N^{-1/2} \sum_i e^{ - i \bd{k} \cdot \bd{r}_i }
 \zeta_i
 \bd{S}_i$, where $\zeta_i = 1$ for $\bd{r}_i \in A$ and
 $\zeta_i = -1$ for $\bd{r}_i \in B$.  For large $S$ we find
to leading order
\begin{subequations}
\begin{eqnarray}
S_{{\rm{st}} , \bd{k}}^x &\approx & N^{-1/2} (S / a^D) \lambda_{\bd{k}} \hat{\Pi}_{ \bd{k} +}, 
\\
S_{{\rm{st}} , \bd{k}}^y &\approx & N^{-1/2} (S / a^D) {\lambda}_{\bd{k}} \hat{\Pi}_{ \bd{k} -},
\end{eqnarray}
\end{subequations}
and
\begin{subequations}
\begin{eqnarray}
S_{ \bd{k}}^x &\approx&  - N^{-1/2}  {\lambda}_{\bd{k}}^{-1} \hat{\Phi}_{ \bd{k} -}, \\
S_{ \bd{k}}^y &\approx&   N^{-1/2}  {\lambda}_{\bd{k}}^{-1} \hat{\Phi}_{ \bd{k} +}.
\end{eqnarray}
\end{subequations}
Here ${\lambda}_{\bd{k}} = [ u_{\bd{k}} + v_{\bd{k}} ]  
( \epsilon_{\bd{k}} /2 )^{1/2} = [(1 + \gamma_{\bd{k}})/2  ]^{1/2}$ 
approaches unity  for ${\bd{k}} \rightarrow 0$.
Hence, our Hermitian field operators can be identified physically with
the suitably normalized transverse components of the staggered and total
(ferromagnetic) spin for  large $S$.  
Writing $\hat{\bd{\Phi}} = ( \Phi_+ , \Phi_- )$ and
$\hat{\bd{\Pi}} = ( \Pi_+ , \Pi_-  )$ and
using Eq.~(\ref{eq:psiplus}), we find
that $\hat{H}_2^{\prime} $ takes the form 
\begin{eqnarray}
  \hat{H}_2^{\prime}  & = & 
\frac{1}{V}\sum_{ \bd{k}  }
  \Bigl[ \frac{
  \hat{\bd{\Phi}}_{ - \bd{k} } \cdot {\hat{ \bd{\Phi}}}_{ \bd{k}  }}{ 2 \chi_0}
 + \frac{\chi_0  E_{ \bd{k}}^2  }{2}
 \hat{\bd{\Pi}}_{ - \bd{k} } \cdot \hat{\bd{\Pi}}_{\bd{k} } 
  \Bigr] \; .
 \label{eq:H2osc}
\end{eqnarray}

The parameterization of the spin fluctuations introduced
by Anderson in his early work
on the antiferromagnetic ground state \cite{Anderson52} 
differs from our 
Eq.~(\ref{eq:psiplus}) only in the choice of
normalization factors.
Subsequently, this method
has been far less popular than the 
approach based on the usual bosonic annihilation and
creation operators 
and has been largely forgotten.
Nevertheless, the Hermition operator approach has many advantages:
(a) 
the suppression of the effective interaction
between antiferromagnetic magnons at long wavelengths is manifest;
(b) the relation between the Holstein-Primakoff bosons and the
continuum fields $\bd{\Pi}$ in the NLSM can be made precise;
(c) a new effective action for the transverse staggered spin 
fluctuations can be obtained 
by eliminating the ferromagnetic fluctuations in a path
integral formulation. The resulting interaction vertices
differ from those of the NLSM, because the 
$1/S$-approach contains scattering processes which are not related to the
constant length constraint of the spin vector.
At one-loop order it turns out that
the two-body  interaction does not renormalize the
Gaussian part of the effective action, so that
the quantum critical point separating the renormalized classical from
the quantum disordered phase can be directly related to the anomalous dimension
of the $\bd{\Pi}$-field;
(d) 
the Hermitian operator approach is most convenient to discuss
spin-waves  in finite-size  antiferromagnets.

We now explain the above points in some detail, beginning with (d).
As first pointed out by 
Anderson \cite{Anderson52},
for finite $N$ the 
$\bd{k} =0$ term 
in  Eq.~(\ref{eq:H2osc}) should be treated with special care.
Defining the dimensionless operators 
$\bd{P} = ( S N )^{-1/2} \hat{\bd{\Phi}}_{ 0 }$ and
$\bd{X} = (S / N )^{1/2} a^{-D} \hat{\bd{\Pi}}_{0 }$ whose components satisfy
$ [ X_{\sigma} , P_{\sigma^{\prime}} ] =i  \delta_{ \sigma, \sigma^{\prime}}$, 
and using the fact that  for periodic boundary conditions  $E_{\bd{k} =0} =0$,
the contribution from the $\bd{k} =0$ term to Eq.~(\ref{eq:H2osc}) 
is
 $\hat{H}_2^0 = \frac{ \bd{P}^2}{ 2 m}$ with $m = 4 D JS$.
Obviously, the spectrum of $\hat{H}_2^0$ is continuous;
the ground state is the zero-momentum state
satisfying $\langle \bd{P}^2 \rangle =0$.
 However, for any state
 with $ \langle \bd{X} \rangle
= \langle \bd{P} \rangle=0$ we have
$ \langle \bd{P}^2  \rangle \langle \bd{X}^2 \rangle \geq 1$ by the uncertainty
principle,  so that
$\langle \bd{X}^2 \rangle = \infty$ in the ground state.
The staggered magnetization  
$M_{\rm st} = \langle \sum_i \zeta_i S^{z}_i \rangle$ can be written as
 \begin{eqnarray}
 M_{\rm st} & = & N ( S + 1/2 ) -
 \langle \bd{P}^2 \rangle  - \langle  \bd{X}^2   \rangle 
- \frac{1}{2V} 
\sum_{ \bd{k} \neq 0}  
 \bigl\langle
f_{\bd{k}}
  \hat{\bd{\Phi}}_{ - \bd{k}} \cdot
\hat{\bd{\Phi}}_{  \bd{k}}  +
 f_{\bd{k}}^{-1}
  \hat{\bd{\Pi}}_{ - \bd{k}} \cdot
\hat{\bd{\Pi}}_{  \bd{k}}
\bigr\rangle
 \; ,
 \label{eq:Mstdef}
\end{eqnarray}
where $f_{\bd{k}} = a^D / ( S \lambda_{\bd{k}}^2 )$. 
In the ground state of $\hat{H}_2$ the
zero mode gives rise to an infinite correction to $M_{\rm st}$.
At first sight, it appears that our approach is inconsistent. 
The reason why the spin-wave approach can still be useful in finite systems
has been discussed by Anderson\cite{Anderson52}: suppose we prepare
a finite-size antiferromagnet by some external field in  a minimal uncertainty wave-packet
with  
$ \langle \bd{P}^2  \rangle  = \langle \bd{X}^2 \rangle = 1$.
The zero-mode contribution to $M_{\rm st}$ is then
of relative order $1/N$, and can be ignored for large $N$.
The validity of the spin-wave approach becomes
then a dynamical problem:
from  quantum mechanics we know 
that under the time evolution governed by 
 $\hat{H}_2^0 = \frac{ \bd{P}^2}{ 2 m}$ the width of the Gaussian wave-packet
increases as $ \langle \bd{X}^2 \rangle_t = 
\langle \bd{X}^2 \rangle_0 + \frac{t^2}{m^2}  \langle \bd{P}^2 \rangle_0$
so that the spin-wave approximation breaks down at 
time   $t \approx  \sqrt{N} m = \sqrt{N} / ( 4 D J S )$, which
for  macroscopic systems exceeds the time scale of experiments.

We now go beyond Ref.~\cite{Anderson52} and consider
interactions between spin-waves.
Within the Holstein-Primakoff transformation, the leading interaction
correction is given by the following bosonic two-body Hamiltonian
\begin{eqnarray}
  \hat{H}_4
  &=&- \frac{1}{8}
  \sum_{ij} J_{ij} \big[ 4 n_i n_j 
  +n_i b_i b_j 
  + b_i n_j b_j   
  + b^{\dagger}_i b^{\dagger}_j n_j + 
  b^{\dagger}_i n_i b^{\dagger}_j \big]
  \label{eq:H4}
  \; ,
\end{eqnarray}
where $n_i = b^{\dagger}_i b_i$.
It is now straightforward (although quite tedious) to express 
$\hat{H}_4$ in terms of our Hermitian field operators.
Since we shall later use the phase space path integral \cite{Schulman81}
to eliminate the ferromagnetic fluctuations, 
we symmetrize $\hat{H}_4$ whenever
the vertices involve non-commuting operators \cite{Gollisch01}.
The final result is
\begin{eqnarray}
 \hat{H}_4   & = & E_4 + \frac{\hat{H}_2^{\prime}}{2S}+
 \frac{ \rho _0  }{ 2  V^3} \frac{D}{a^2} 
 \sum_{\bd{k}_1,\ldots, \bd{k}_4}
  \delta_{ \bd{k}_1+\bd{k}_2+\bd{k}_3+\bd{k}_4,0}
   \Bigl[
   \Gamma^{\Pi\Pi} ( \bd{k}_1 , \bd{k}_2 , \bd{k}_3 , \bd{k}_4 )
   \hat{\bd{\Pi}}_{ \bd{k}_1} \cdot \hat{\bd{\Pi}}_{ \bd{k}_2}  
   \hat{\bd{\Pi}}_{ \bd{k}_3} \cdot
   \hat{\bd{\Pi}}_{ \bd{k}_4}
 \nonumber
 \\
  & &\hspace{4cm} +
   {f_0^2} 
   \Gamma^{\Phi \Pi}_{\parallel}
   ( \bd{k}_1 , \bd{k}_2 , \bd{k}_3 , \bd{k}_4 )
   \sum_{\sigma}  \frac{1}{2} \big\{
   \hat{\Phi}_{ \bd{k}_1 \sigma}  \hat{\Phi}_{ \bd{k}_2 \sigma},
   \hat{\Pi}_{ \bd{k}_3 \sigma}  \hat{\Pi}_{ \bd{k}_4 \sigma} \big\}
\nonumber
\\ &&\hspace{4cm}
   +
  f_0^2 \Gamma^{\Phi \Pi}_{\perp} 
   ( \bd{k}_1 , \bd{k}_2 , \bd{k}_3 , \bd{k}_4 )
  \sum_{\sigma}
   \hat{\Phi}_{ \bd{k}_1 \sigma}  \hat{\Phi}_{ \bd{k}_2 \sigma}  
   \hat{\Pi}_{ \bd{k}_3, - \sigma}  \hat{\Pi}_{ \bd{k}_4, -\sigma}
   \nonumber
   \\
   & &   \hspace{4cm}+
  f_0^2 \Gamma^{ \Phi \Pi}_{u}
   ( \bd{k}_1 , \bd{k}_2 , \bd{k}_3 , \bd{k}_4 )
    \big\{
   \hat{\Phi}_{ \bd{k}_1 +}  , \hat{\Pi}_{ \bd{k}_2 +}
   \big\}
   \big\{
   \hat{\Pi}_{ \bd{k}_3 -}  , \hat{\Phi}_{ \bd{k}_4 -}
   \big\}
\nonumber
\\ &&\hspace{4cm}
   +f_0^4\Gamma^{\Phi \Phi } ( \bd{k}_1 , \bd{k}_2 , \bd{k}_3 , \bd{k}_4 )
   \hat{\bd{\Phi}}_{ \bd{k}_1} \cdot \hat{\bd{\Phi}}_{ \bd{k}_2}  
   \hat{\bd{\Phi}}_{ \bd{k}_3} \cdot
   \hat{\bd{\Phi}}_{ \bd{k}_4} \Bigr] \; ,
  \label{eq:H4trans}
\end{eqnarray}
where $\{\hat{A},\hat{B} \}=\hat{A}\hat{B}+\hat{B}\hat{A}$ 
denotes the anticommutator,
$E_4=-  ( D  J/4)  [ N + \sum_{\bd k} \gamma_{\bd k}] $, and $f_0= f_{ \bd{k} =0} = a^D/S$. 
Writing $\gamma_{\bd{1}} = \gamma_{ \bd{k}_1 }$ etc.,
the properly symmetrized vertices in Eq.~(\ref{eq:H4trans}) are
\begin{subequations}
    \begin{eqnarray}
      \Gamma^{\Pi \Pi} ( \bd{k}_1 , \bd{k}_2 , \bd{k}_3 , 
      \bd{k}_4 ) & = &
       (1/8)
       \lambda_{ \bd{1} }   \lambda_{ \bd{2}  }   
      \lambda_{ \bd{3}  }   \lambda_{ \bd{4} }  
      \bigl[ \gamma_{\bd{1}} + \gamma_{\bd{2}} + 
      \gamma_{\bd{3}} + \gamma_{\bd{4}} -
      2 ( \gamma_{ \bd{1} + \bd{2}}  +  \gamma_{ \bd{3} + \bd{4}}) \bigr]
      \; ,
      \label{eq:GammaPiPi}
      \\
      \Gamma_{\parallel}^{\Phi \Pi} ( \bd{k}_1 , \bd{k}_2 , \bd{k}_3 , 
      \bd{k}_4 ) & = &
    (1/4)  
    ( \lambda_{ \bd{1} }   \lambda_{ \bd{2} } )^{-1}   
       \lambda_{ \bd{3} }   
      \lambda_{ \bd{4} }  
       \bigl[    
      - \gamma_{\bd{1}} - \gamma_{\bd{2}} 
      +  \gamma_{ \bd{3}}  +  \gamma_{\bd{4}}    
      - 2 ( \gamma_{\bd{1} + \bd{2} } +
       \gamma_{\bd{3} + \bd{4}} ) 
      \bigr]
      \; ,
      \label{eq:GammaPhiPipar}
      \\
      \Gamma_{\perp}^{\Phi \Pi}( \bd{k}_1 , \bd{k}_2 , \bd{k}_3 , 
      \bd{k}_4 ) & = &
     (1/4)
     ( \lambda_{ \bd{1} }   \lambda_{ \bd{2} } )^{-1}   
      \lambda_{ \bd{3} }   
      \lambda_{ \bd{4} }
      \bigl[   
      3 ( 
      - \gamma_{\bd{1}} - \gamma_{\bd{2}} 
      +  \gamma_{ \bd{3}}  +  \gamma_{\bd{4}} )  
      - 2 ( \gamma_{\bd{1} + \bd{2} }  + \gamma_{\bd{3} + \bd{4} } ) 
\nonumber \\
&& \hspace{3.1cm}+ 4 ( \gamma_{\bd{1} + \bd{3} } +
         \gamma_{\bd{2} + \bd{4}} ) 
      \bigr]
      \; ,
      \label{eq:GammaPhiPiperp}
      \\
      \Gamma^{\Phi  \Pi}_{u} ( \bd{k}_1 , \bd{k}_2 , \bd{k}_3 , 
      \bd{k}_4 ) & = &
      (1/8)
      \lambda_{ \bd{1} }^{-1}   \lambda_{ \bd{2} }   
      \lambda_{ \bd{3} }   
      \lambda_{ \bd{4} }^{-1}
      \bigl[
      \gamma_{\bd{1}} - \gamma_{\bd{2}} 
      -  \gamma_{ \bd{3}}  +  \gamma_{\bd{4}} -  
      4 ( \gamma_{\bd{1} + \bd{3} } +  \gamma_{\bd{2} + \bd{4}} )
      \bigr]
      \label{eq:GammaPhiPiu}
      \; ,
      \\
      \Gamma^{\Phi \Phi} ( \bd{k}_1 , \bd{k}_2 , \bd{k}_3 , 
      \bd{k}_4 ) & = &
         (1/8)
     ( \lambda_{ \bd{1} }   \lambda_{ \bd{2}  }   
      \lambda_{ \bd{3}  }   \lambda_{ \bd{4} } )^{-1}  
      \bigl[ - \gamma_{\bd{1}} - \gamma_{\bd{2}} - 
      \gamma_{\bd{3}} - \gamma_{\bd{4}} -
      2 ( \gamma_{ \bd{1} + \bd{2}}  +  \gamma_{ \bd{3} + \bd{4}} )   
      \bigr]
      \; .
      \label{eq:GammaPhiPhi}
    \end{eqnarray}
\end{subequations}
The important point is now that the
vertex $\Gamma^{\Pi \Pi} $ associated with the 
direct interaction between staggered spin fluctuations
vanishes quadratically for small momenta, 
\begin{eqnarray} 
  \Gamma^{\Pi \Pi} ( \bd{k}_1 , \bd{k}_2 , \bd{k}_3 , 
      \bd{k}_4 ) &=&  
( a^2 / 16 D) 
[\bd{k}_1^2+\bd{k}_2^2+\bd{k}_3^2+
\bd{k}_4^2 
+4\bd{k}_1\cdot\bd{k}_2+4\bd{k}_3\cdot\bd{k}_4]
    +{\cal O}(\bd{k}_i^4) \; , 
 \label{eq:vertexpi}
 \end{eqnarray}
while the other vertices 
approach finite limits,
\begin{equation}  
\Gamma^{\Phi \Pi}_{\parallel} (0) = \Gamma^{\Phi \Pi}_{u} (0)=  
\Gamma^{\Phi \Phi} (0) =  -1\, , \, \, \mbox{and} \, \,
\Gamma^{\Phi \Pi}_{\perp} (0) = 1\, .
\label{eq:vertexlimit}
\end{equation}
Hence, in our parameterization the bare interaction between
the staggered spin fluctuations
is manifestly suppressed at 
long wavelengths, as seen from Eq.~(\ref{eq:vertexpi}).
General symmetry arguments \cite{Hofmann99} suggest that this suppression 
survives when we take into account the renormalization of the
staggered spin fluctuations by the
ferromagnetic ones.
We have explicitly verified this within the Hartree-Fock approximation,
where the two-body part in Eq.~(\ref{eq:H4trans}) is replaced by
a one-body Hamiltonian.   Using
$\langle \hat{\Pi}_{ - \bd{k} \sigma } \hat{\Pi}_{  \bd{k} \sigma }
 \rangle = V ( 2 \chi_0 E_{\bd{k}} )^{-1}$,
$\langle \hat{\Phi}_{ - \bd{k} \sigma } \hat{\Phi}_{  \bd{k} \sigma }
 \rangle = V  \chi_0 E_{\bd{k}} /2 $, and
$\langle \hat{\Pi}_{ - \bd{k} \sigma } \hat{\Phi}_{  \bd{k} \sigma }
 \rangle = i V /2 $, we find
that the  Hartree-Fock approximation amounts to the
replacement $\hat{H}_4 \rightarrow E_4^{\prime} + \frac{C}{2S} \hat{H}_2^{\prime}$, where $E_4^{\prime}$ differs from $E_4$ given above by a
constant of the order of unity, and 
$C = 1 - (2 /N) \sum_{\bd{k}}
 \epsilon_{\bd{k}}$. 
It follows that the leading $1/S$-correction to the 
magnon dispersion can be taken into account in 
the free spin-wave Hamiltonian $\hat{H}_2^{\prime}$ given
in Eq.~(\ref{eq:H2osc}) by  replacing
$\chi_0  \rightarrow Z_{\chi} \chi_0$ and $
E_{\bd{k}} \rightarrow Z_c E_{\bd{k}}$, with the
renormalization factors $Z_{\chi} = [ 1 + \frac{C}{2S} ]^{-1}$
and $Z_c =    1 + \frac{C}{2S} $.
Note that this simply
leads to an overall rescaling of  the magnon dispersion, but
does not change its wave-vector dependence \cite{Chinn71}.

In contrast, the vertices encountered
in the usual spin-wave theory, based on 
the magnon operators $\alpha_{\bd{k}}$ and $\beta_{\bd{k}}$
defined in Eq.~(\ref{eq:bogoliubov}),
have a more complicated structure.
Using the Dyson-Maleev transformation (the vertices derived from
the Holstein-Primakoff transformation are linear combinations
of the Dyson-Maleev vertices \cite{Harris71}),
the interaction part
takes the form \cite{Kopietz97,Castilla91}
\begin{eqnarray}
  \hat{H}_4^{\rm DM}&=&\frac{\chi_0^{-1}}{4V}
  \sum_{{\bd k}_1,\dots , {\bd k}_4} \Big\{
  V^{(1)}_{1234}(\beta_1^\dagger \beta_2^\dagger\beta_3\beta_4+
  \alpha_3^\dagger \alpha_4^\dagger \alpha_1 \alpha_2)-
  2V^{(2)}_{1234}(\alpha_3^\dagger \beta_4 \alpha_1 \alpha_2+
  \alpha_4^\dagger \beta_1^\dagger \beta_2^\dagger \beta_3)
  \nonumber \\
  && \hspace{1.9cm}
  +2 V^{(3)}_{1234}( \alpha_3^\dagger \alpha_4^\dagger \alpha_2 \beta_1^\dagger
  + \alpha_1 \beta_2^\dagger \beta_3 \beta_4)
  -2 V^{(4)}_{1234}(\alpha_3^\dagger \alpha_1 \beta_2^\dagger \beta_4
  +\alpha_4^\dagger \alpha_2 \beta_1^\dagger \beta_3)\nonumber \\
  &&\hspace{1.9cm}
  +V^{(5)}_{1234}(\alpha_3^\dagger\alpha_4^\dagger\beta_1^\dagger
  \beta_2^\dagger +\alpha_1\alpha_2\beta_3
  \beta_4 \Big\} \delta_{\bd{k}_1+\bd{k}_2,\bd{k}_3+\bd{k}_4} \, .
\end{eqnarray}
For simplicty, we have introduced the notation $\alpha_i=\alpha_{\bd{k}_i}$
etc.
In the limit that all wave vectors are small, the 
Dyson-Maleev vertices 
$ V^{(j)}_{1234}=V^{(j)}({\bd k}_1,{\bd k}_2,{\bd k}_3,{\bd k}_4)$
behave as
\cite{Kopietz97}
\begin{eqnarray}
  V^{(j)}_{1234}
  \sim \frac{1}{2}
  \sqrt{\frac{|{\bd k}_1| |{\bd k}_2|}{|{\bd k}_3| |{\bd k}_4|}}
  \left(1 +\xi_j \frac{\bd{k}_1 \cdot \bd{k}_2}
    {|{\bd k}_1| |{\bd k}_
      2|} \right) \, ,
  \label{eq:DMvertices}
\end{eqnarray}
with $\xi_1=\xi_2=\xi_5=1$ and $\xi_3=\xi_4=-1$. 
Obviously, these vertices do not vanish for small momenta.
Moreover, the long-wavelength
limits are direction-dependent and the non-analytic prefactor 
can potentially give rise to divergences in perturbation theory. 
Although the weakness of the underlying spin-wave interaction is not
aparent,
it was shown in \cite{Kopietz90} that to order $1/S^2$ the
divergences cancel in a $1/S$-expansion as a consequence
of total spin conservation \cite{Maleev00}. 
Comparing Eq.~(\ref{eq:DMvertices}) to the equivalent 
Eqs.~(\ref{eq:vertexpi},\ref{eq:vertexlimit}) of the Hermitian operator
formulation, the much simpler structure of the latter becomes clear.
This simpler structure is a direct consequence of using physically
transparent operators.

In the conventional approach based on the 
operators $\alpha_{\bd{k}}$ and $\beta_{\bd{k}}$,
it is very cumbersome to calculate
higher order terms in a
$1/S$-expansion \cite{Harris71,Kopietz90}.
Instead of performing a complete $1/S$-expansion, we
shall here only concentrate on the long wavelength
antiferromagnetic fluctuations. 
Within the 
Hermitian field parameterization these are clearly
separated from the ferromagnetic ones.
Moreover, the
weakness of the interaction between long wavelength staggered fluctuations
is explicit in this approach.
We thus simply eliminate the  ferromagnetic degrees of freedom
represented by the $\hat{\bd{\Phi}}$-operators 
and work directly with the effective action
$ S_{\rm eff} [ \bd{\Pi} ]$ of the staggered fluctuations.
Formally, $ S_{\rm eff} [ \bd{\Pi} ]$ can be defined 
as a phase space path-integral \cite{Schulman81,Gollisch01},
$ e^{-S_{\rm eff} [ \bd{\Pi} ] } = \int {\cal{D}} [ \bd{\Phi}  ]
 e^{- S [ \bd{\Pi} , \bd{\Phi} ]}$,
where the Euclidean action
$ S [ \bd{\Pi}, \bd{\Phi}  ]$ is a functional of
quantum fields  $\bd{\Pi}_{\bd{k}} ( \tau )$ and
$\bd{\Phi}_{\bd{k}} ( \tau )$ 
associated with the field operators.
Within the Gaussian approximation we obtain
\begin{equation}
  S  [ \bd{\Pi} , \bd{\Phi} ]  \approx 
  \frac{1}{2} 
  \int_K  \Bigl[   \chi_0^{-1} \bd{\Phi}_{ - K } \cdot \bd{\Phi}_{K} 
 + \chi_0 E_{\bd{k}}^2  \bd{\Pi}_{ - K } \cdot \bd{\Pi}_{K} 
    -   \omega_n  ( \bd{\Phi}_{ - K } \cdot \bd{\Pi}_{K } 
  - \bd{\Pi}_{ - K } \cdot \bd{\Phi}_{K } ) 
  \Bigr]
  \; ,
  \label{eq:S2pp}
\end{equation}
where $K=({\bd k},i\omega_n)$ is a collective label for
wave-vector and bosonic Matsubara frequency $\omega_n$,
the Fourier transformed fields are 
$\bd{\Pi}_{K } = \int_0^{\beta} d \tau e^{ i \omega_n \tau} \bd{\Pi}_{\bd{k}} ( \tau )$, and
$\int_K=(\beta V)^{-1}\sum_{{\bd k} \omega_n}$.
The corresponding Gaussian effective action for the staggered fluctuations is
 \begin{equation} 
S_{\rm eff} [ \bd{\Pi} ]  \approx
\frac{ \chi_0}{2} \int_K  ( E_{\bd{k}}^2 + \omega_n^2 )
 \bd{\Pi}_{ -K } \cdot \bd{\Pi}_{ K }
 \; .
 \label{eq:Seffgauss}
 \end{equation}
In the long-wavelength limit
$E_{\bd k}\approx c_0 | \bd{k} |$, so that
Eq.~(\ref{eq:Seffgauss}) 
has precisely the same
form as the Gaussian approximation of the 
NLSM  in Eq.~(\ref{eq:sigma}).
Because we have explicitly constructed the $\bd{\Pi}$-field 
in Eq.~(\ref{eq:Seffgauss})  from the Holstein-Primakoff bosons
$b_i$, we have succeeded to give a mathematically precise relation
between the two different parameterization of the spin fluctuations.

The  leading interaction correction to 
$S_{\rm eff}   [ \bd{\Pi} ]$ is due to
the vertex $\Gamma^{\Pi \Pi}$ in  Eq.~(\ref{eq:H4trans}) 
involving four $\hat{\bd{\Pi}}$-operators.
Retaining only this term 
we obtain  in the long-wavelength limit  with the help of
Eq.~(\ref{eq:vertexpi}) at $T=0$,
\begin{eqnarray}
  S_{\rm eff} [ \bd{ \Pi} ] &   \approx &  
  \frac{ 1 }{2g_0} \int d x_0   \int d^{D} x 
  \Bigl\{
( \partial_{\mu} \bd{\Pi})^2 
 + ( \partial_{0}  \bd{\Pi})^2
+ \frac{1}{4}  \bigl[ 2\left({\bd{\Pi}} \cdot\partial_{\mu}{\bd{\Pi}}\right)^2 
-{\bd{\Pi}}^2 
 (\partial_{\mu}{\bd{\Pi}} )^2  \bigr] \Bigr\}
  \; ,
  \label{eq:Sint4}
\end{eqnarray}
where $\bd{x} = \Lambda_0 \bd{r}$ and $x_0 = \Lambda_0 c_0 \tau$
are dimensionless space-time variables,
$g_0 = \Lambda_0^{D-1} c_0 / \rho_0$ is 
the usual \cite{Chakravarty89} 
dimensionless coupling constant of the NLSM, and
$\Lambda_0 \approx 1/a$ is an ultraviolet cutoff. 
The quartic interaction
in Eq.~(\ref{eq:Sint4})
does {\it{not}} agree with the 
quartic term 
in the perturbative expansion of the action (\ref{eq:sigma}) of the
NLSM~\cite{Chakravarty89}. In the latter case,
the term in the square braces of Eq.~(\ref{eq:Sint4}) is replaced by
$ 4 ({\bd{\Pi}} \cdot \partial_{\mu}{\bd{\Pi}})^2$ and there is an additional
counter-term due to the expansion of the functional $\delta$-function
enforcing the constraint $\bd{\Omega}^2 =1$.
The reason for this difference is that 
in the NLSM all
interactions arise from the 
non-linear constraint $\bd{\Omega}^2 =1$. In contrast,  the
$1/S$-approach contains dynamic interactions neglected in the NLSM, but 
ignores the constraints associated with the finiteness of the spin Hilbert space
which according to Dyson \cite{Dyson56} amount to irrelevant
kinematic interactions.

It turns out that  within a  one-loop momentum shell renormalization 
group (RG) approach \cite{Kopietz89,Chakravarty89}
the quartic interaction in Eq.~(\ref{eq:Sint4})
does not renormalize the Gaussian part (\ref{eq:Seffgauss}) of our effective action
at all. A similar cancellation is known to happen also
in the corresponding ferromagnetic Heisenberg model \cite{Kopietz89}.
If we apply the usual momentum scale RG procedure
consisting of mode elimination and rescaling \cite{Kopietz89,Chakravarty89},
then the dependence of the running coupling $g_l$
on the logarithmic RG flow parameter $l$ is at one loop entirely due to the
rescaling step.
The resulting one-loop RG flow equation is simply
\begin{equation}
\partial_l g_l=- (D-1-\eta^{\pi}_l) g_l \, .
\end{equation} Here
$\eta_l^{\pi}$ is the running anomalous dimension
of the $\bd{\Pi}$-field (not to be confused with the usual critical exponent $\eta$), 
which is related to the interaction-dependent part of the field rescaling factor $Z_l^{1/2}$ as usual,
$\eta_l^{\pi} = - \partial_l \ln Z_l$.
The simplest way to calculate  $\eta^{\pi}_l$ is from the
transformation properties of the staggered magnetization 
$M_{\rm st}$ given in Eq.~(\ref{eq:Mstdef})
under the momentum scale RG \cite{Kopietz89}.
By demanding that the scaling of the fluctuation correction
in Eq.~(\ref{eq:Mstdef}) is consistent with the leading term 
$N ( S + 1/2)$ (which scales as the volume), it is easy to show that
$\eta^{\pi}_l = \frac{K_D}{2} g_l$, where $K_D$ is the surface of 
the unit sphere in $D$ dimensions divided by $(2 \pi )^D$.
The resulting RG equation for $g_l$, 
\begin{equation}
\partial_l g_l=(1-D) g_l+\frac{1}{2} K_D \, g_l^2 \, ,
\end{equation}
is of course identical to the
equation derived in Ref.\cite{Chakravarty89}.
Interestingly, in our formulation the
quantum critical point separating the renormalized classical from the
quantum disordered regime can be  characterized by
$\lim_{ l \rightarrow \infty} \eta_l^{\pi} = D-1$.

In summary, we have developed a new parameterization of the
spin-wave expansion which is clearly superior to
the conventional formulation based on the
magnon operators $\alpha_{\bd{k}}$ and $\beta_{\bd{k}}$ in
Eq.~(\ref{eq:bogoliubov}). In our parameterization  the physical meaning
of the degrees of freedom is transparent, the weak interaction between
long-wavelength staggered magnons is manifest, the subtleties
of the spin-wave approach in finite-size systems can be easily discussed, and
the effective theory for the staggered fluctuations can be derived.
With our method it is also straightforward to discuss
the effect of interactions between the zero modes on
the finite-size spin-wave spectrum, and
to study spin-wave interactions in
more complicated 
models involving various anisotropies or external magnetic fields \cite{Hasselmann05}.

\acknowledgements
We thank Florian Sch\"{u}tz und Ivan Spremo for useful discussions and appreciate 
financial support by the DFG.

\end{document}